# Interface Doping Mechanism of p-Type Conductivity in Mist-CVD-Grown $LiGa_5O_8$ Thin Films

Dong Su Yu[1], Shashu Tomar[2], Zixiu Huang[1], Kaitian Zhang[1], Sarker Md. Sadman[3], Vijay Gopal Thirupakuzi Vangipuram[1], Ye Lin[1], Walter R. L. Lambrecht[3], Roberto C. Myers[1,2,4] and Hongping Zhao[1,4,*]

[1]*Department of Electrical and Computer Engineering, The Ohio State University, Columbus, OH 43210, USA*

[2]*Department of Physics, The Ohio State University, Columbus, OH 43210, USA*

[3]*Department of Physics, Case Western Reserve University, Cleveland, OH 43106, USA*

*4Department of Materials Science and Engineering, The Ohio State University, Columbus, OH 43210, USA*

[*]Corresponding author Email: zhao.2592@osu.edu

## Abstract

$LiGa_5O_8$ is an ultrawide-bandgap oxide semiconductor that has recently been experimentally demonstrated to exhibit p-type conductivity; however, the origin of the observed p-type transport remains poorly understood. In this work, we investigate the origin of p-type conductivity by systematically comparing the electrical, compositional, and structural properties of mist chemical vapor deposition (mist-CVD) grown $LiGa_5O_8$ thin films exhibiting either p-type conductivity or insulating behavior under different growth conditions. The p-conductive films exhibit room-temperature hole concentrations on the order of $10^{18}$ $cm^{-3}$. The emergence of p-type conductivity is found to be strongly correlated with the elemental composition of the films, with X-ray photoelectron spectroscopy (XPS) revealing a reduced Li/Ga ratio in the p-conductive films, indicative of a Li-deficient composition. Raman spectroscopy and X-ray diffraction (XRD) further reveal signatures of $\beta$-$Ga_2O_3$ in the p-conductive films that are absent in the insulating films, demonstrating the formation of a secondary $\beta$-$Ga_2O_3$ phase. Considering the reported type-II band alignment between $LiGa_5O_8$ and β-$Ga_2O_3$, together with theoretically predicted acceptor levels associated Li vacancies ($V_{Li}$) in $LiGa_5O_8$, the formation of mixed $LiGa_5O_8$/β-$Ga_2O_3$ phases is proposed to facilitate electron transfer from the β-$Ga_2O_3$ valence band to acceptor states in $LiGa_5O_8$, thereby promoting hole accumulation and p-type conductivity.

## I. Introduction

Ultrawide-bandgap (UWBG) oxide semiconductors are increasingly studied for high-power electronic devices[1-4] and ultraviolet optoelectronic devices[5-7]. However, achieving reliable p-type conductivity in UWBG oxides remains challenging. In β-$Ga_2O_3$, the relatively flat valence band results in a large hole effective mass, while strong hole self-trapping leads the formation of localized small polarons, making p-type conductivity difficult to achieve.[8-10] To address this limitation, several p-type materials, including NiO[11-15] and $Cr_2O_3$[16-18], have been investigated for use with β-$Ga_2O_3$ in p–n heterojunction devices. More recently, $LiGa_5O_8$ has emerged as a promising p-type UWBG oxide, exhibiting a bandgap wider than that of β-$Ga_2O_3$.[19-22]

Zhang et al.[22] reported room-temperature p-type conductivity in spinel $LiGa_5O_8$ thin films grown by mist chemical vapor deposition (mist-CVD), reporting an optical bandgap of approximately 5.36 eV and hole concentrations up to $10^{18}$ $cm^{-3}$ with a maximum room-temperature hole mobility of 20 $cm^2/V\cdot s$. Temperature-dependent Hall measurements confirmed hole conduction down to 150 K. These findings are notable for such a wide bandgap oxide and indicate that $LiGa_5O_8$ may serve as a promising p-type material for UWBG electronic devices.

The origin of this p-type conductivity remains unclear. First-principles calculations indicate that common acceptor defects in $LiGa_5O_8$ are too deep to generate significant free hole concentrations at room temperature.[23-28] Lyons[26] found that holes strongly localize in $LiGa_5O_8$ and that native acceptors, such as Li vacancies ($V_{Li}$), Ga vacancies ($V_{Ga}$), and Li-on-Ga antisites ($Li_{Ga}$), form deep acceptor levels. Dabsamut et al.[27,28] reached similar conclusions, showing that $V_{Li}$ are too deep to explain the observed hole concentrations and that donor defects such as oxygen vacancies ($V_O$) and Ga-on-Li antisites ($Ga_{Li}$) can compensate acceptors and keep the Fermi level away from the valence-band maximum. These results suggest that the observed p-type conductivity may happen from a mechanism that does not exist in pure $LiGa_5O_8$. In this paper we provide evidence that $\beta$-$Ga_2O_3$ occurs as a secondary phase and propose an interface doping mechanism of holes from $LiGa_5O_8$ into the $\beta$-$Ga_2O_3$ regions as the source of p-type doping.

Mixed phase formation can occur during Ga-based oxide thin films growth depending on the local composition and growth conditions. Such mixed phase formation has been reported in several Ga-based oxide materials, including AlGaO and spinel gallates, demonstrating that multiple crystalline phases can coexist within a single thin film.[29-33]

The electronic structure at the $LiGa_5O_8$/β-$Ga_2O_3$ hetero-interface offers a key insight. Zhang[34] et al. extracted the band alignment between mist-CVD-grown $LiGa_5O_8$ and β-$Ga_2O_3$ using X-ray photoelectron spectroscopy (XPS), finding a type-II band alignment with a valence-band offset of 1.86 ± 0.09 eV. The $LiGa_5O_8$ valence-band maximum is below that of β-$Ga_2O_3$. Particularly, the deep acceptor levels calculated for $LiGa_5O_8$ are located near or below this energy. Li vacancies which are the lowest energy of formation acceptor type defect in $LiGa_5O_8$ have been calculated to lie at 1.25 eV [26] or 1.56 eV [28] and while deep in $LiGa_5O_8$, lie below the VBM of $Ga_2O_3$ at the interface and thus could readily inject holes into the latter.

## II. Experimental Section

To better understand the differences between p-type conductive and insulating $LiGa_5O_8$ films, detailed compositional and structural characterization was performed on $LiGa_5O_8$ thin films grown by mist-CVD under different growth conditions. All films were grown on 6° offcut c-plane sapphire substrates. The precursor solution was prepared by dissolving gallium acetylacetonate and lithium acetoacetate in a mixture of deionized (DI) water and hydrochloric acid (HCl). The growth temperature was varied from 800 to 950 °C, while the $O_2$ flow rate was set to either 0.1 or 0.2 L/min. The growth duration was 60 or 120 min. The detailed growth conditions for each sample are summarized in Table I.

XPS measurements were performed using a Thermo Fisher Scientific Nexsa G2 XPS system to determine the relative compositions of Li, Ga, and O in the films. Raman spectroscopy was performed using a Renishaw inVia confocal Raman microscope to detect Raman-active vibrational modes and identify the corresponding crystalline phases. The Raman spectra were processed using RamanSPy [35], and the peaks were fitted using Imfit Python package. X-ray diffraction (XRD) measurements were performed using a Bruker D8 Discover diffractometer to examine the crystal structure and phase composition. Room-temperature Hall measurements were performed using the van der Pauw method with an Ecopia HMS-3000 Hall measurement system to determine the carrier type, carrier concentration, and Hall mobility. Calculations of the Raman spectrum were done using Density functional perturbation theory as implemented in the ABINIT code. [25]

## III. Results and Discussions

Room-temperature Hall measurements were used to compare the electrical properties of the mist-CVD-grown $LiGa_5O_8$ films. As summarized in Table I, samples #1 and #2 exhibited p-type conductivity, with hole concentrations of $1.5 \times 10^{18}$ and $1.8 \times 10^{18}$ $cm^{-3}$ and Hall mobilities of 2.1 and 1.9 $cm^2/V \cdot s$, respectively. Samples #3 and #4 were electrically insulating under the same Hall measurement conditions. These samples were used to compare the electrical properties of $LiGa_5O_8$ films grown under different growth conditions. Samples #1 and #4 were selected as representative p-conductive and insulating films, respectively, for detailed compositional and structural characterization. The clear difference between the conductive and insulating films suggests that the growth conditions strongly influence the electrical properties of $LiGa_5O_8$, although the origin of this difference cannot be determined from the Hall measurements alone.

The compositions of the p-conductive and insulating $LiGa_5O_8$ films were compared by XPS. Figure 1 shows the Li 1s, Ga 3s, and O 1s spectra of the p-conductive (sample #1) and insulating (sample #4) films. The atomic ratios obtained from the XPS measurements are summarized in Table II. The p-conductive film shows lower Li/Ga and Li/O ratios of 0.163 and 0.116, respectively, compared with 0.219 and 0.190 for the insulating film. The ideal stoichiometric ratios for $LiGa_5O_8$ are Li/Ga = 0.20 and Li/O = 0.125. These results indicate that the p-type conductive thin film is relatively Li-poor, while the insulating film appears relatively L-rich, particularly based on the Li/O ratios. This compositional difference is associated with the observed difference in electrical conductivity.

Raman spectroscopy was used to further compare the crystalline phases of the p-conductive and insulating $LiGa_5O_8$ films. Figure 2 shows the Raman spectra of the two $LiGa_5O_8$ films together with reference spectra measured from β-$Ga_2O_3$, c-plane sapphire, and $LiGaO_2$. Both $LiGa_5O_8$ films were grown on 6° offcut c-plane sapphire substrates, a sapphire reference spectrum was scanned to distinguish substrate-related Raman peaks from those associated with the films. The p-conductive and insulating films exhibit Raman spectrum characteristic of the $LiGa_5O_8$ phase.[25,36-37] However, several additional peaks are observed in the p-conductive $LiGa_5O_8$ film. In particular,

additional Raman peaks at 167.9±0.3, 199.5±0.3, 318.9±0.6, 346.6±0.4, 414.8±0.5, 474.3±0.6, 627±1, 658.4±0.6, and 766.4±0.5 $cm^{-1}$ are observed in the p-conductive film but are not observed in the insulating $LiGa_5O_8$ film. These peak positions are consistent with Raman modes observed in the β-$Ga_2O_3$ reference spectrum. In contrast, the characteristic Raman peaks of $LiGaO_2$ do not match the additional peaks observed in the p-conductive $LiGa_5O_8$ film. The peaks at 378.6±0.4, 417.1±0.4, 429.9±0.5, 448.6±0.3, 576.8±0.3, and 750.2±0.3 $cm^{-1}$ are detected in both p-conductive and insulating $LiGa_5O_8$ films as well as in the 6° offcut c-plane sapphire reference spectrum, indicating that these peaks are associated with the substrate. In contrast, the additional peaks identified in the p-conductive film are not observed in the sapphire reference spectrum. These results therefore support the coexistence of $LiGa_5O_8$ and β-$Ga_2O_3$ phase in the p-conductive film, while no clear β-$Ga_2O_3$-related peaks are observed in the insulating $LiGa_5O_8$ film.

Fig. 3 shows the conducting (b) and insulating $LiGa_5O_8$ (c) spectra after subtracting the substrate peaks and compared again with β-$Ga_2O_3$ in (a) and with the calculated spectrum of $LiGa_5O_8$ in (d). The latter is calculated using Density Functional Perturbation Theory using the GGA-PBE functional [25], which tends to overestimate lattice constants and hence underestimate bond strengths, and thus gives peaks slightly red-shifted from the experiment. We here show the crystal orientation and polarization averaged spectrum because the measurements did not use polarization filters. For non-polarization filtered measurements on a (111) oriented film, all Raman active $A_1$, E and $T_2$ modes are observable and the peaks in the calculated spectrum are correspondingly labeled [25]. Corresponding peaks in the different panels are indicated by dashed lines. Note that the strongest $A_1$ peak in the calculated spectrum at 403 $cm^{-1}$, which is the third $A_1$ mode, occurs at 424 $cm^{-1}$ in experiment [36-37] and overlaps with the sapphire substrate peak. Hence the subtraction procedure may underestimate the $LiGa_5O_8$ contribution to this peak.

To further investigate the structural differences between the p-conductive and insulating $LiGa_5O_8$ films, 2θ–ω XRD measurements were performed with a resolution of 0.01°. Figure 4 presents a comparison of the XRD peaks for both films across a broad 2θ range from 10° to 90°. Both samples exhibit primary diffraction peaks corresponding to crystalline $LiGa_5O_8$, along with a pronounced (0006) peak from the $Al_2O_3$ substrate. The $LiGa_5O_8$ (111), (222), (333), and (444) peaks are observed at approximately 18.92°, 38.07°, 58.50°, and 81.21°, respectively. While the overall diffraction patterns appear similar, differences between the two samples become more apparent upon higher resolution scans of specific $LiGa_5O_8$ peaks.

Figure 5 shows high resolution 2θ–ω XRD scans around the (222) and (333) $LiGa_5O_8$ peaks collected with a resolution of 0.001°. In both regions, the p-conductive film shows additional diffraction intensity on the higher-angle side of the main $LiGa_5O_8$ peaks. In the p-conductive film (sample #1), additional peaks are observed at approximately 38.365° and 59.155°, corresponding to the β-$Ga_2O_3$ ($40\bar{2}$) and ($60\bar{3}$) planes, respectively. In contrast, these β-$Ga_2O_3$ peaks are not observed in the insulating $LiGa_5O_8$ film (sample #4). Together with the Raman results shown in Figs. 2 and 3, these additional peaks support the presence of a secondary β-$Ga_2O_3$ phase in the p-conductive film. In contrast, the insulating film is primarily composed of the $LiGa_5O_8$ phase, with no clear evidence of β-$Ga_2O_3$. This difference implies that the phase composition of the films may influence their electrical properties.

The observation of a mixed β-$Ga_2O_3$ phase in the p-conductive $LiGa_5O_8$ film is consistent with the composition-dependent p-type conductivity reported previously by Zhang et al[22]. In that work, stronger p-type conductivity was associated with Li-poor, Ga-rich, and O-poor compositions. The presence of β-$Ga_2O_3$ in the present p-conductive film is therefore consistent with a locally Li-deficient and Ga-rich composition relative to stoichiometric $LiGa_5O_8$. This agreement suggests that the mixed phase formation observed here can be closely related to the compositional ratios associated with p-type conductivity.

In order to explain the mechanism of hole generation and transport in $LiGa_5O_8$ with mixed β-$Ga_2O_3$ phases, Fig. 6 plots the band alignment between $LiGa_5O_8$ and β-$Ga_2O_3$. As reported from our previous studies[19], a type-II band alignment is formed between $LiGa_5O_8$ and β-$Ga_2O_3$ interface, with a valence-band offset of 1.86 ± 0.09 eV and the $LiGa_5O_8$ valence-band maximum located below that of β-$Ga_2O_3$. First-principles calculations have predicted several native acceptor levels in $LiGa_5O_8$, including those associated with $V_{Li}$ and $Li_{Ga}$. Lyons[26] reported a $V_{Li}$ (0/−) transition level at 1.25 eV and a $Li_{Ga}$ (0/−) transition level at 1.64 eV above the $LiGa_5O_8$ valence-band maximum. Dabsamut et al.[28] also reported a $V_{Li}$ as an acceptor with a transition level at 1.58 eV above the $LiGa_5O_8$ valence-band maximum, while the $Li_{Ga}$ levels more than 2 eV above the VBM. The $V_{Li}$ calculated acceptor level is located below the β-$Ga_2O_3$ valence-band maximum when the measured band offset is considered.

This energy band alignment suggests an interfacial charge-transfer process. Electrons from β-$Ga_2O_3$ valence-band states could transfer into available $Li_{Ga}$- and $V_{Li}$-related acceptor levels in $LiGa_5O_8$, leaving holes accumulated in the β-$Ga_2O_3$ regions. The resulting holes can facilitate hole transport through the film and contribute to the observed p-type conductivity in the conductive $LiGa_5O_8$ samples. Although conventional p-type doping of β-$Ga_2O_3$ is severely limited by the formation of self-trapped polarons, the hole concentration within the β-$Ga_2O_3$ regions can be substantially higher than the average Hall concentration measured for the entire film, particularly when β-$Ga_2O_3$ constitutes only a fraction of the film volume.. This model provides a plausible explanation for the observed correlation between mixed $LiGa_5O_8$/β-$Ga_2O_3$ formation and p-type transport. However, direct experimental evidence for interfacial charge transfer, visualization of the $Ga_2O_3$ regions and their interconnectivity to provide a percolating hole conduction path will require further investigation.

Such an interface charge transfer doping model has previously been proposed for P doped ZnO involving $Zn_3P_2$ as a secondary phase[38]. The band bending near an interface of $LiGa_5O_8$ with a concentration of $10^{19}$-$10^{20}$ $cm^{-3}$ $V_{Li}$ acceptor levels and a band offset of 1.86 eV $\beta$-$Ga_2O_3$ can be modeled using Silvaco Atlas and predicts the formation of a high density hole gas at the VBM of the $Ga_2O_3$ interface region. Figure 7 plots the band diagram and equilibrium Fermi level at the $LiGa_5O_8$/$\beta$-$Ga_2O_3$ heterointerface with the acceptor concentration of $10^{19}$ $cm^{-3}$. A sheet of hole concentration of ~ $3.3x10^{12}$ $cm^{-2}$ is formed at the interface. While $\beta$-$Ga_2O_3$ itself is known to be notoriously difficult to dope p-type because of the high valence and maximum mass and hence self-trapped polaron formation, the high density of holes may screen the Coulomb interaction responsible for the self-trapped polaron formation. Thus, our model, rather than predicting $LiGa_5O_8$ itself to be p-type predicts the very efficient interface modulation doping of $Ga_2O_3$ at interfaces with Li-poor $LiGa_5O_8$. To optimally use this effect, heterostructure design could be employed.

## IV. Conclusion

In this work, the compositional, structural, and electrical properties of p-conductive and insulating mist-CVD-grown $LiGa_5O_8$ thin films were compared to determine the origin of the recently discovered p-type conductivity. Room-temperature Hall measurements revealed p-type conduction with hole concentrations on the order of $10^{18}$ $cm^{-3}$ in selected films, while other films remained insulating. XPS showed a reduced Li/Ga ratio in the p-conductive films, indicating a Li-deficient composition. Raman spectroscopy and XRD further confirmed β-$Ga_2O_3$-related features exclusively in the p-conductive films, indicating the formation of a secondary β-$Ga_2O_3$ phase.

These results establish a strong correlation between p-type conductivity, Li deficiency, and the formation of mixed $LiGa_5O_8$/β-$Ga_2O_3$ phases.

Based on the reported type-II band alignment, in which the β-$Ga_2O_3$ valence-band maximum is approximately 1.86 eV higher than that of $LiGa_5O_8$, and the calculated $Li_{Ga}$- and $V_{Li}$-related acceptor levels in $LiGa_5O_8$, electron transfer from β-$Ga_2O_3$ valence-band states to acceptor levels in $LiGa_5O_8$ is proposed. Such hetero-interfacial charge transfer could generate large concentrations of holes in the β-$Ga_2O_3$ regions and facilitate hole transport within the mixed-phase structure, contributing to the observed p-type Hall response. While $\beta$-$Ga_2O_3$ itself has been found to be notoriously difficult to p-type dope because of self-trapped polaron formation, this may have been limited by low concentrations of holes, which here is overcome by a very efficient interface doping mechanism. These results suggest a possible mechanism linking Li deficiency, β-$Ga_2O_3$ secondary-phase formation, interfacial charge transfer, and p-type conductivity in $LiGa_5O_8$ thin films.

**Acknowledgements**

The authors thank the financial support from the Air Force Office of Scientific Research FA9550-23-1-0142, FA9550-26-1-0005 (AFOSR, Dr. Ali Sayir).

**Data Availability**

The data that support the findings of this study are available from the corresponding author upon reasonable request.

## Table Captions

**Table 1.** Summary of the growth conditions, film thicknesses, and electrical properties of a series of mist-CVD-grown $LiGa_5O_8$ thin films on 6° offcut c-plane sapphire substrates.

**Table 2.** Li/Ga, Li/O, and Ga/O atomic ratios obtained from XPS measurements for the p-conductive (sample #1) and insulating (sample #4) $LiGa_5O_8$ thin films.

## Figure Captions

**Figure 1.** High-resolution XPS spectra of the Li 1s, Ga 3s, and O 1s core levels for the p-conductive (sample #1) and insulating (sample #4) $LiGa_5O_8$ thin films. The spectra were used to compare the relative Li, Ga, and O compositions between the two films.

**Figure 2.** Raman spectra for p-conductive $LiGa_5O_8$ (sample #1), insulating $LiGa_5O_8$ (sample #4), β-$Ga_2O_3$, c-sapphire and $LiGaO_2$ samples.

**Figure 3.** Raman spectrum of conductive (b) and insulating (c) $LiGa_5O_8$ spectra after subtraction of sapphire substrate spectrum, compared with (a) $\beta$-$Ga_2O_3$ Raman spectrum and (d) Calculated $LiGa_5O_8$ spectrum averaged over polarization directions.

**Figure 4.** 2θ-ω XRD scan of p-conductive (sample #1) and insulating (sample #4) $LiGa_5O_8$ thin films grown on 6° offcut c-plane sapphire substrates. The diffraction peaks allocated to spinel cubic $LiGa_5O_8$ and the (0006) $Al_2O_3$ substrate reflection are indicated.

**Figure 5.** High resolution 2θ-ω XRD scans of the p-conductive (sample #1) and insulating (sample #4) $LiGa_5O_8$ thin films within the 2θ ranges of (a) 37.5–39.0°; and (b) 57.5–60.0°.

**Figure 6.** Schematic band alignment and proposed interfacial charge-transfer mechanism in mixed-phase $LiGa_5O_8$/β-$Ga_2O_3$. Electrons are transferred from the valence band of β-$Ga_2O_3$ to $Li_{Ga}$- and $V_{Li}$- related defect states in $LiGa_5O_8$, resulting in hole accumulation in the β-$Ga_2O_3$ region.

**Figure 7. (**Bottom) Band diagram for a $LiGa_5O_8$/β-$Ga_2O_3$ interface. (Top) Corresponding density of holes.

**Table 1**

| **Sample #** | **$O_2$ flow (L/min)** | **Growth temperature (°C)** | **Growth duration (min)** | **Film thickness (nm)** | **Hole concentration ($cm^{-3}$)** | **Hall mobility ($cm^2/V \cdot s$)** |
|---|---|---|---|---|---|---|
| **1** | 0.1 | 900 | 120 | 575.6 | $1.5 \times 10^{18}$ | 2.1 |
| **2** | 0.1 | 950 | 120 | 606.4 | $1.8 \times 10^{18}$ | 1.9 |
| **3** | 0.1 | 800 | 60 | 208.4 | Insulating | |
| **4** | 0.2 | 850 | 120 | 405.8 | Insulating | |

**Table 2**

| Sample | Li/Ga | Li/O | Ga/O |
|---|---|---|---|
| p-conductive $LiGa_5O_8$ (sample #1) | 0.163 | 0.116 | 0.711 |
| Insulating $LiGa_5O_8$ (sample #4) | 0.219 | 0.190 | 0.865 |

**Figure 1**

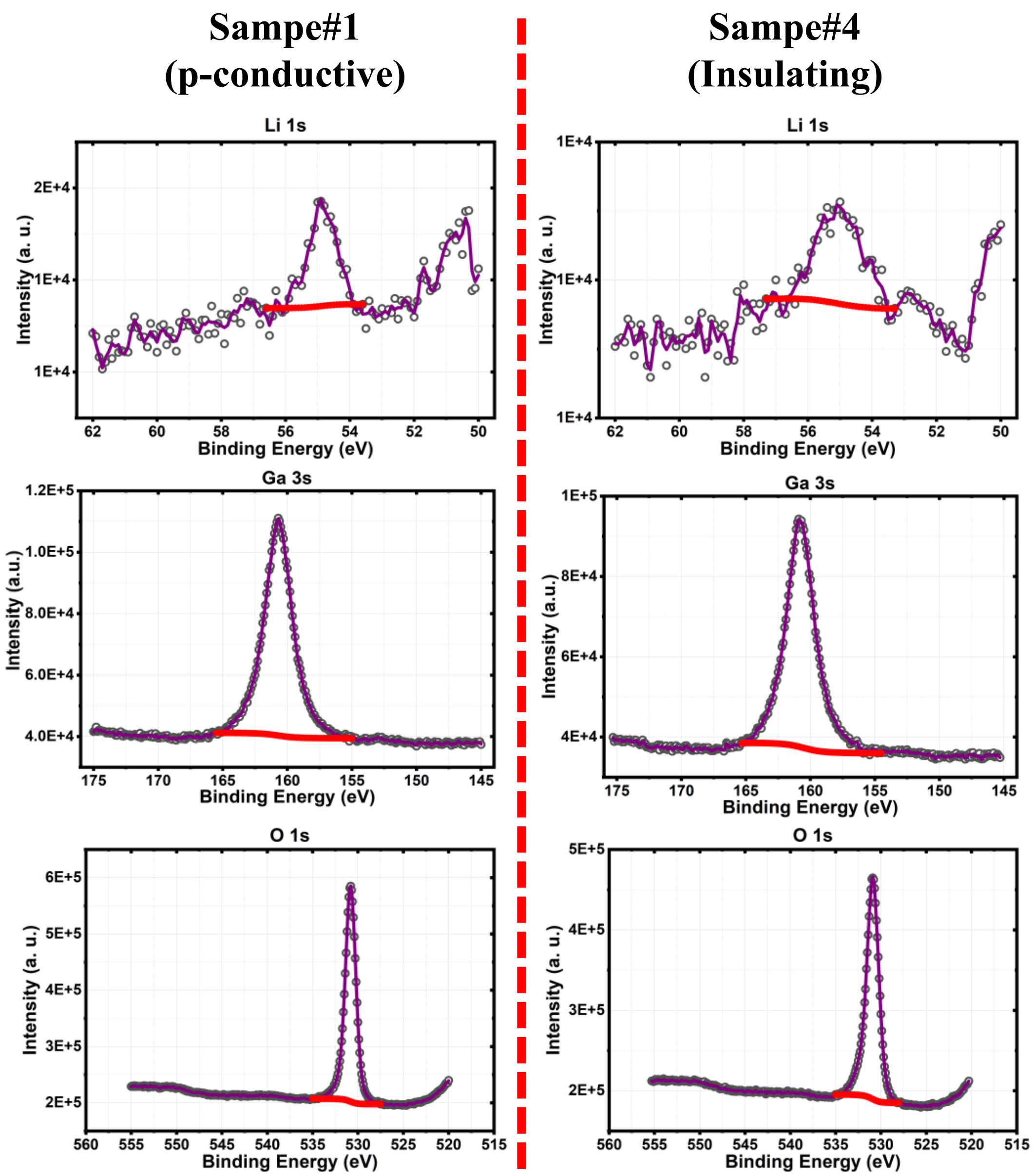

Sampe#1
(p-conductive)
Sampe#4
(Insulating)
Li 1s
Ga 3s
O 1s
Intensity (a. u.)
Intensity (a.u.)
Binding Energy (eV)

**Figure 2**

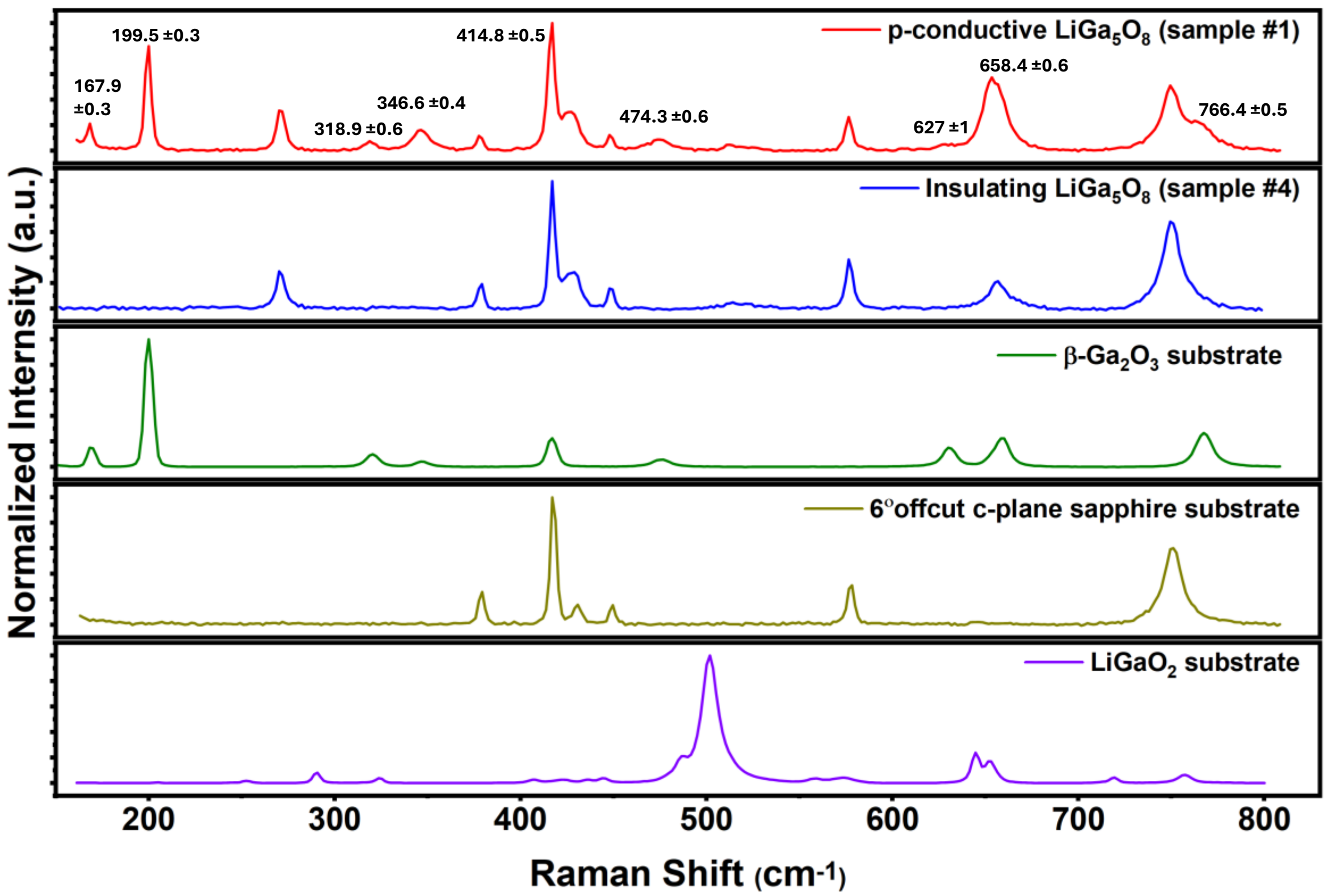

p-conductive LiGa5O8 (sample #1)
Insulating LiGa5O8 (sample #4)
β-Ga2O3 substrate
6° offcut c-plane sapphire substrate
LiGaO2 substrate
167.9 ±0.3
199.5 ±0.3
318.9 ±0.6
346.6 ±0.4
414.8 ±0.5
474.3 ±0.6
627 ±1
658.4 ±0.6
766.4 ±0.5
Normalized Intensity (a.u.)
Raman Shift (cm-1)
200
300
400
500
600
700
800

**Figure 3**

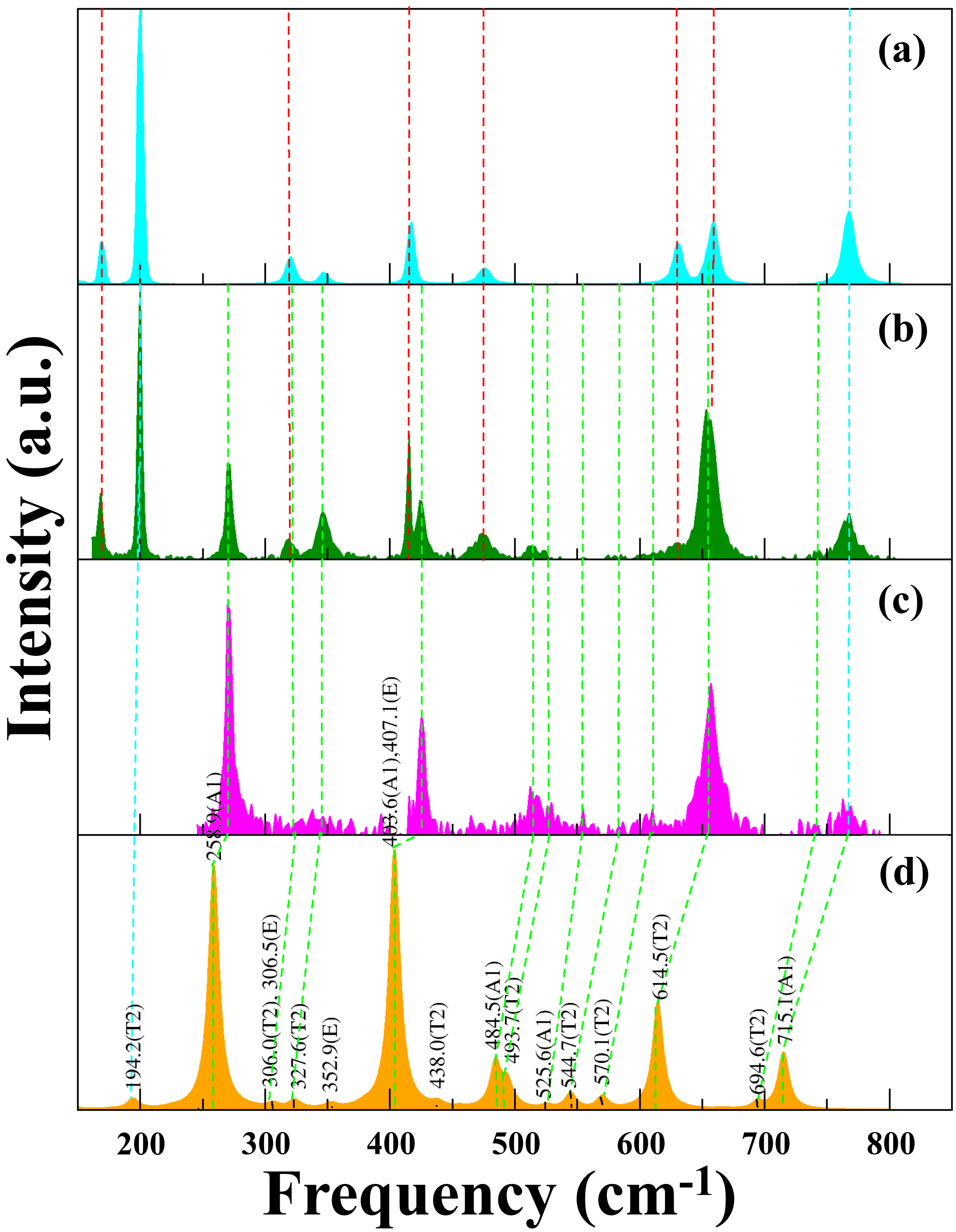

(a)
(b)
(c)
(d)
Intensity (a.u.)
Frequency (cm-1)
200
300
400
500
600
700
800
194.2(T2)
258.9(A1)
306.0(T2), 306.5(E)
327.6(T2)
352.9(E)
403.6(A1),407.1(E)
438.0(T2)
484.5(A1)
493.7(T2)
525.6(A1)
544.7(T2)
570.1(T2)
614.5(T2)
694.6(T2)
715.1(A1)

**Figure 4**

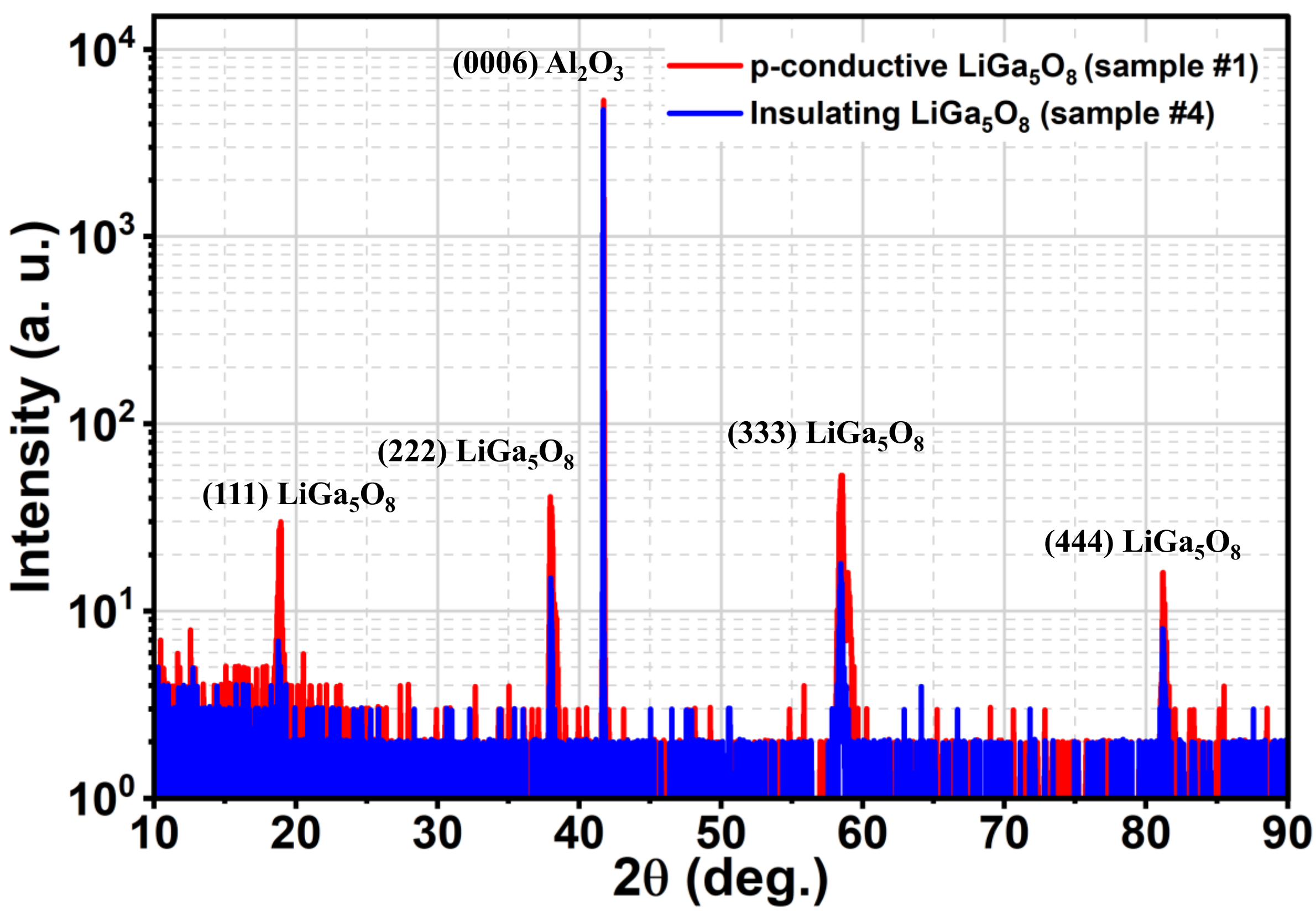

(0006) $Al_2O_3$
p-conductive $LiGa_5O_8$ (sample #1)
Insulating $LiGa_5O_8$ (sample #4)
(111) $LiGa_5O_8$
(222) $LiGa_5O_8$
(333) $LiGa_5O_8$
(444) $LiGa_5O_8$
Intensity (a. u.)
$10^0$
$10^1$
$10^2$
$10^3$
$10^4$
10 20 30 40 50 60 70 80 90
2θ (deg.)

**Figure 5**

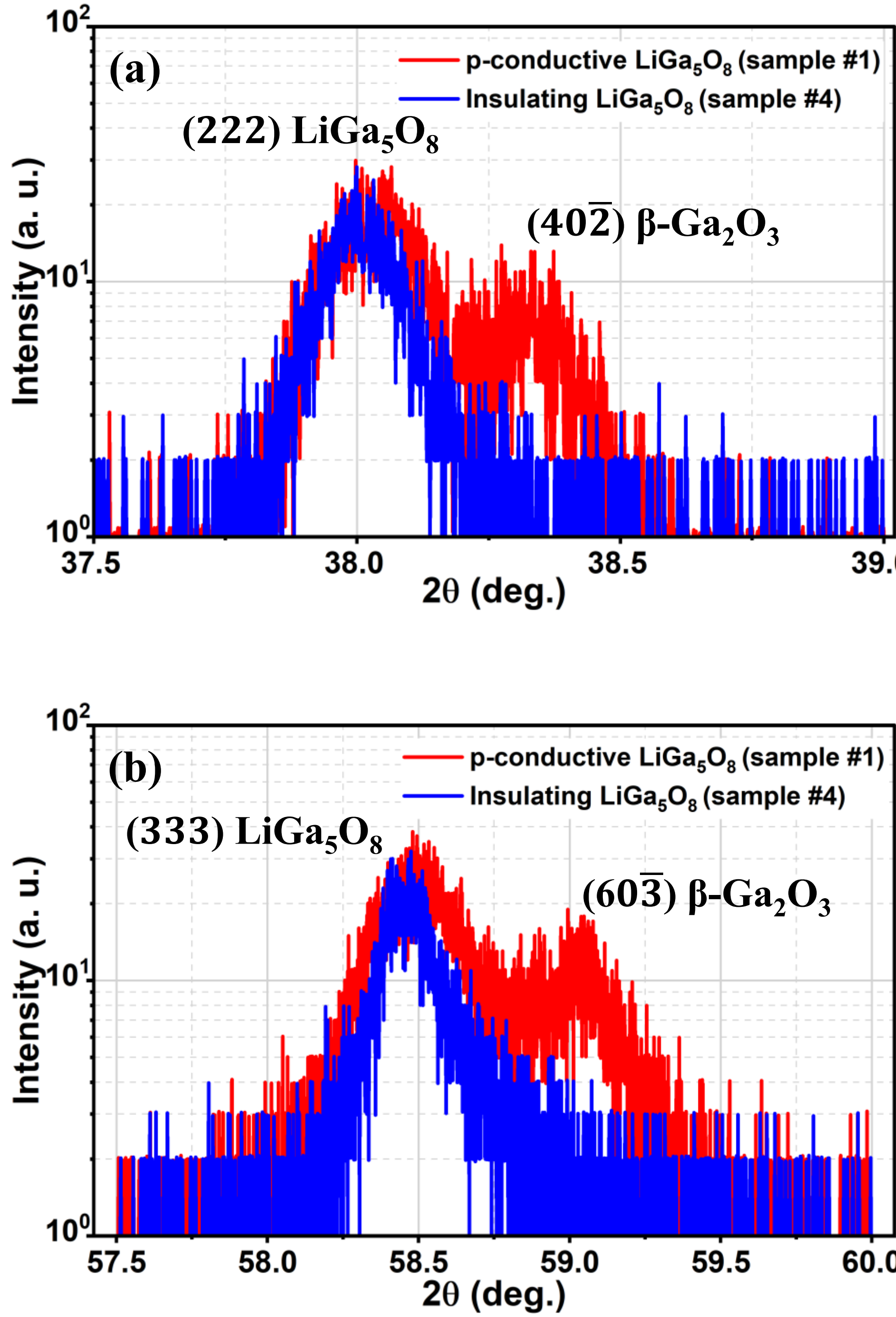

(a)
p-conductive LiGa5O8 (sample #1)
Insulating LiGa5O8 (sample #4)
(222) LiGa5O8
(40$\overline{2}$) β-Ga2O3
Intensity (a. u.)
2θ (deg.)
(b)
p-conductive LiGa5O8 (sample #1)
Insulating LiGa5O8 (sample #4)
(333) LiGa5O8
(60$\overline{3}$) β-Ga2O3
Intensity (a. u.)
2θ (deg.)

**Figure 6**

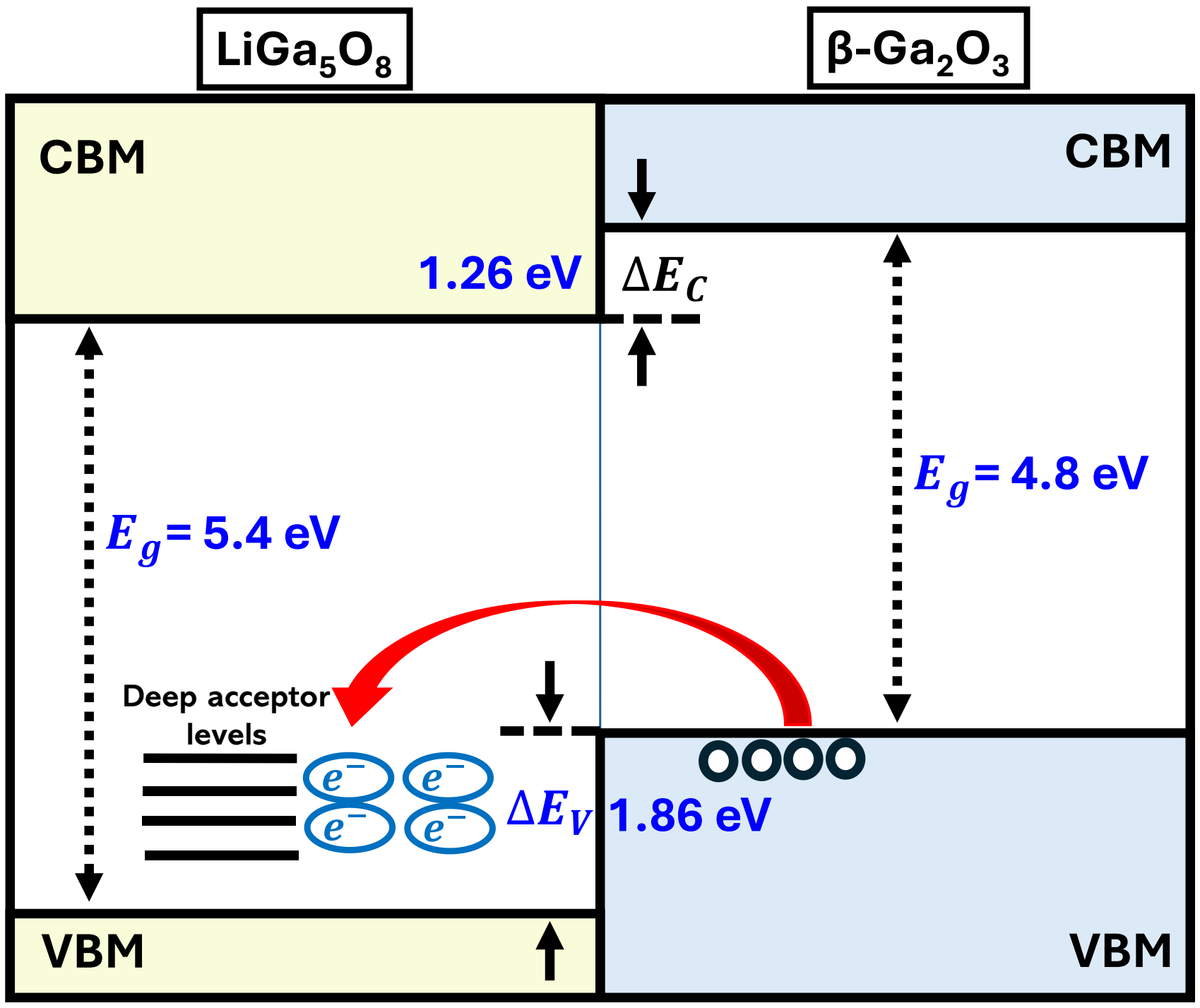

LiGa5O8
β-Ga2O3
CBM
CBM
1.26 eV
ΔEC
Eg = 5.4 eV
Eg = 4.8 eV
Deep acceptor
levels
e−
e−
e−
e−
ΔEV
1.86 eV
VBM
VBM

**Figure 7**

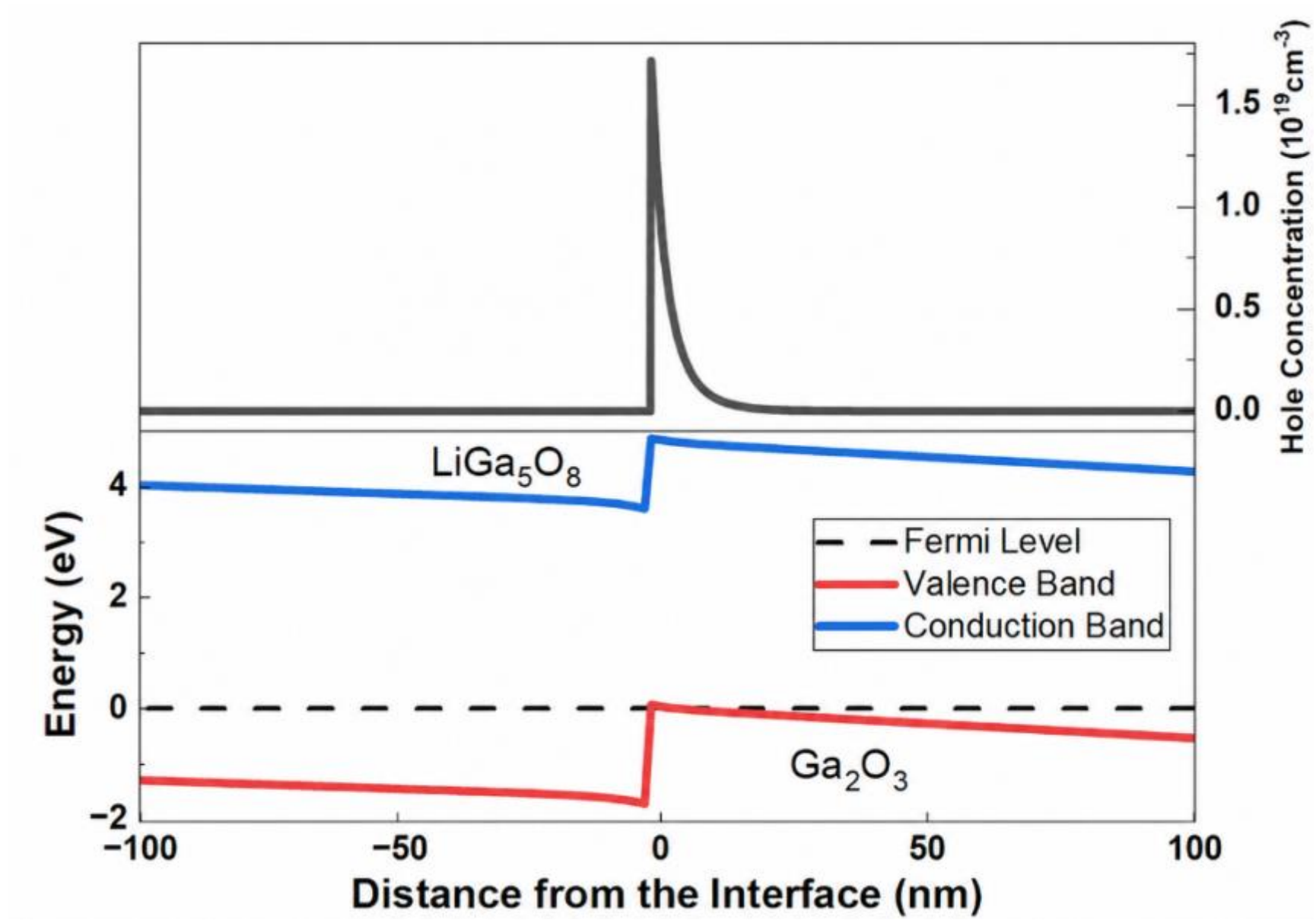

Hole Concentration ($10^{19}cm^{-3}$)
1.5
1.0
0.5
0.0
$LiGa_5O_8$
Energy (eV)
4
2
0
-2
Fermi Level
Valence Band
Conduction Band
$Ga_2O_3$
-100
-50
0
50
100
Distance from the Interface (nm)